\documentstyle[aps,twocolumn]{revtex}
\input{psfig}

\begin{document}
\draft
\def\be{\begin{equation}}
\def\ee{\end{equation}}
\def\beq{\begin{eqnarray}}
\def\eeq{\end{eqnarray}}
\def\fractwo{\frac{1}{2}}
\wideabs{
\title{
Composite Fermions with Spin Freedom
}
\author
{
Daijiro {\sc Yoshioka}
}
\address{Department of Basic Science, The University of Tokyo\\
3-8-1 Komaba, Meguro-ku, Tokyo 153-8902, Japan
}
\date{\today}
\maketitle

\begin{abstract}
General rule for the composite fermion transformation, when the spins of the
electrons are not polarized is derived.
Condition for the quantum phase transition between various spin states is
obtained based on the rule.
This rule gives foundation for the experimental determination of the mass and
$g$-factor of the composite fermion.
\end{abstract}

\pacs{Keywords:
fractional quantum Hall effect, composite fermion, g-factor
}
}


%
%
\section{Introduction}
\begin{figure}
\psfig{figure=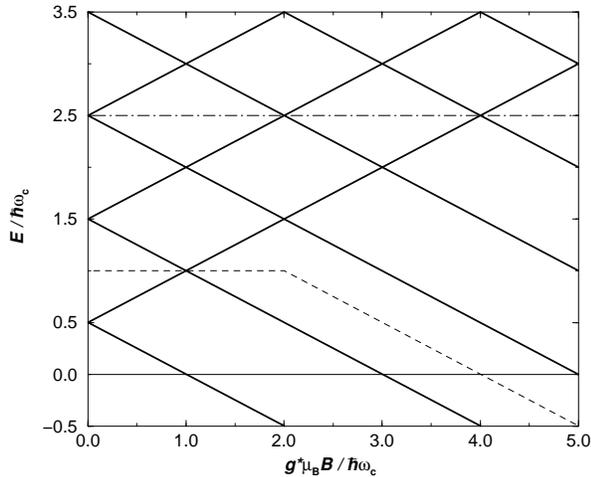,height=7cm}
\caption{Energy of the Landau levels as functions of the Zeeman splitting,
$g^*\mu_{\rm B}B$.
The level crossing occurs when the horizontal axis takes an integer value $j$.
The dashed line and the dash-dotted line are the Fermi level at $T=0$ for
$\nu=2$ and 5, respectively.}
\label{fig:1}
\end{figure}

Energy gap due to the Landau quantization or the spin Zeeman splitting is
essential for the integer quantum Hall effect.\cite{klitz,prange}
When the Landau level filling factor $\nu$ is equal to an integer $N$, the
lowest $N$ spin-split Landau levels are filled, and the IQHE will be observed,
in principle.
However, if we can change the Landau level spacing $\hbar \omega_{\rm c}$ and
the size of the Zeeman splitting $g^*\mu_{\rm B}B$ independently, level
crossings of the Landau levels belonging to the different spin states can be
caused.
Here $\omega_{\rm c}$ is the cyclotron frequency, $g^*$ is the $g$-factor of
the two-dimensional electron and $\mu_{\rm B}$ is the Bohr magneton.
As depicted in Fig.~\ref{fig:1}, the spin splitting of the Landau levels is
proportional to $g^*\mu_{\rm B}B$, and the condition for the level crossing is
given as
\be
j\hbar\omega_{\rm c} = g^* \mu_{\rm B}B,
\ee
where $j$ is an integer.

%
%
If the level crossing occurs for the Landau levels at the Fermi energy,
configuration of the electrons changes.
This is a quantum phase transition, and the energy gap vanishes at the
transition.
As shown in Fig.~\ref{fig:1} the phase transition occurs only when $j=1$ for
$\nu=2$, while it occurs at $j=2$ and 4 for $\nu=5$.

Since, the fractional quantum Hall effect can be understood as an integer
quantum Hall effect of the composite fermions (CF),\cite{jain,heinonen} similar
quantum phase transition between different spin polarization is expected to
occur.
Actually, such transitions are observed experimentally,\cite{du,yeh} and they
have been used to deduce the effective mass $m^*$ and $g$-factor of the
composite fermion $g^*$.
In these experiments tilting of the magnetic field is used to enhance the spin
Zeeman splitting.

The interpretation of the experiment is quite simple, as long as the filling
factor is expressed as $\nu=p/(mp+1)$, or its electron-hole conjugate, $\nu =
2-p/(mp+1)$, where $p= \pm 1, \pm 2,...$ is an integer and $m$ is an even
integer.
Here $|p|$ gives the filling factor of the composite fermion and $m$ describes
the number of flux quanta attached to the electron to transform it to a
composite fermion.
On the other hand, for the filling factor such as $\nu=5/7$ or 4/5, it is not
simple.
For example let us consider the situation at $\nu=5/7$.
According to Wu {\it et al.},\cite{wu,jain1} $\nu=(3p+2)/(4p+3)$ is related to
$\nu'=(3p+2)/(2p+1)$ by a transformation, $\nu=\nu'/(2\nu'-1)$, which in turn
is related to $\nu''=p/(2p+1)$ by the electron-hole symmetry.
This state finally maps into $\nu'''=p$.
Therefore $\nu=5/7$, which is obtained by inserting $p=1$ to the above series,
is mapped to $\nu'''=1$.
If this mapping is appropriate, there should be no quantum phase transition,
since there is no level crossing for the lowest Landau level.
However, another mapping is possible.
If the Zeeman splitting is quite large, $\nu=5/7$ can be considered as an
electron-hole symmetric state of $\nu=2/7$.
This filling has the form of $\nu=p/(mp+1)$ with $p=-2$, and $m=4$.
In this case, since the filling factor of the composite fermion is 2, quantum
phase transition between the spin-polarized ground state and spin-singlet
ground state seems possible.
However, as the true electron-hole symmetric state of $\nu=5/7$ is not
$\nu=2/7$ but $\nu=1+2/7$, we first need to establish a rule to handle the
situation where both spin states are occupied before we can consider the
possibility of the transition.

Now the experiment clearly shows that there is a quantum phase transition at
$\nu=5/7$.\cite{yeh}
Therefore, Wu {\it et al.}'s interpretation is not appropriate.
Furthermore, we cannot simply treat $\nu=5/7$ as an electron-hole symmetric
state of $\nu=2/7$.
This is what Yeh {\it et al.} did in ref.(\cite{yeh}).
In their treatment, they could not relate the observed collapse of excitation
gap to level crossing {\it at the Fermi level}.
Namely, they tried to determine $m^*$ and $g^*$ so that energy gap collapse is
related to the level crossing of the Landau levels.
Then they found that a unique choice of the product $m^*g^*$ can relate every
feature in the resistivity to level crossings in a consistent way.
However, then the most prominent feature in the resistivity at $\nu=5/7$ had to
be connected to the condition that twice of the Landau level splitting is equal
to the Zeeman splitting, i.e. it occurs at $j=2$.
At this point the filling of the lowest two Landau levels will not change: the
level crossing occurs between the third and fourth lowest Landau levels as
shown in Fig.~\ref{fig:1}.
Therefore, $\nu=5/7$ should not be mapped to $\nu=2/7$, for which the CF
filling factor is two.
The situation was the same for other filling factors around $\nu=3/4$: None of
the prominent features in the resistivity could be related to the quantum phase
transition, if states around $\nu=3/4$ are considered as electron-hole
symmetric states around $\nu=1/4$.

I have briefly pointed out that the apparent discrepancy is resolved, if we
correctly take into account both spin states of the original
electrons.\cite{dy}
It is pointed out that the spin freedom of the composite fermion should be
understood as a result of that of the electrons.
In the present paper, I give details of the theory.

The organization of this paper is as follows.
In \S 2 we consider the FQH states at $\nu=p/(2p+1)$ and its electron-hole
symmetric states at $\nu=2-p/(2p+1)$.
The latter states can be considered as those at $\nu=1+p'/(2p'+1)$.
{}From this analysis we establish a set of rules for the composite fermion
transformation for $\nu=1 \pm p/(mp+1)$.
These rules are applied to the states at $m=4$, or at $\nu=1 \pm p/(4p+1)$ in
\S 3.
We derive a condition for the quantum phase transition in these filling
factors.
Comparison with the experiments and discussion are given in \S 4.

%
%
\section{Electron-Hole Symmetry}
In this section we consider states at
\be
\nu=\frac{p}{(2p+1)},
\ee
and whose electron-hole symmetric states to deduce a set of rules for the
states at
\be
\nu=1+\frac{p'}{(2p'+1)},
\ee
where $p$ and $p'$ are integers.
In this paper we consider systems in the strong magnetic field limit.
Namely, we consider the electronic cyclotron energy, $\hbar\omega_c$, to be
infinitely large, although we assume that the $g$-factor is small so that the
spin Zeeman splitting is finite.
Therefore, we retain only the lowest Landau levels for each spin state.
In this situation, the electron-hole symmetric states of that at
\be
\nu= \frac{p}{(2p+1)},
\ee
is realized at
\be
\nu'=2 - \frac{p}{(2p+1)}.
\ee

When the Zeeman splitting is large enough, the states at $\nu=p/(2p+1)$ is
spin-polarized, and mapped to composite fermion states at $\nu_{\rm CF}=|p|$.
In the experiments the Zeeman splitting is enhanced by tilting the magnetic
field:
The spin-Zeeman splitting is proportional to the total magnetic field, $B_{\rm
tot}$, and given by $g^*\mu_{\rm B} B_{\rm tot}$.
On the other hand the Landau level splitting of the CF is given by $\hbar
\omega_{\rm c}^*$, where
\be
\omega_{\rm c}^* = \frac{eB_{\rm eff}}{m^*},
\ee
$B_{\rm eff}$ is the effective magnetic field for the CF, and $m^*$ is the
effective mass of the CF.
The effective field is proportional to the component of the magnetic field
perpendicular to the 2-d plane, $B_{\perp}$:
\be
B_{\rm eff} = \frac{B_\perp}{2p+1} = B_\perp - B_{\perp,1/2}.
\ee
Here $B_{\perp,1/2}$ is the magnetic field to realize the half-filled state.
When the Zeeman splitting is reduced while keeping the size of the Landau level
splitting fixed, successive quantum phase transitions to partially
spin-polarized states occur as stated in the introduction, and observed
experimentally.
The condition for the transition is given as follows.
We denote the filling factor of the up(down)-spin CF as $\nu_{{\rm CF}\uparrow
(\downarrow)}$.
We consider the transition from
\be
\{
 \begin{array}{rl}
   \nu_{{\rm CF}\downarrow} &= |p|-k \\
   \nu_{{\rm CF}\uparrow}   &= k, \\
 \end{array}
\ee
to
\be
 \{
  \begin{array}{rl}
   \nu_{{\rm CF}\downarrow} &= |p|-k-1 \\
   \nu_{{\rm CF}\uparrow}   &= k+1, \\
 \end{array}
\ee
where $k$ is a non-negative integer.
In the state before the transition the energy of the highest occupied Landau
level (HOLL) of down-spin CF is
\be
E_\downarrow =(|p|-k-\fractwo)\hbar\omega_{\rm c}^*
- \fractwo g^*\mu_{\rm B}B_{\rm tot},
\ee
and that of the lowest unoccupied Landau level (LULL) of up-spin CF is
\be
E_\uparrow =(k+\fractwo)\hbar\omega_{\rm c}^* + \fractwo
g^*\mu_{\rm B}B_{\rm tot}.
\ee
At the transition these energies coincide, and the excitation energy vanishes.
The condition is therefore given as
\be
(|p|-2k-1)\hbar\omega_{\rm c}^* =g^*\mu_{\rm B}B_{\rm tot}
\label{cond1}
\ee
This condition has been used in the experiments to deduce the $g$-factor and
effective mass of the CF.
Notice that this condition is not identical to the level crossing condition,
$j\hbar\omega_{\rm c}^* =g^*\mu_{\rm B}B_{\rm tot}$.
The range and parity of $j$ are fixed in eq.(\ref{cond1}).
This restriction is important.

Since, the state at $\nu'=2-p/(2p+1)$ is the electron-hole symmetric state of
that at $\nu=p/(2p+1)$,
the same condition, eq.(\ref{cond1}), should be satisfied at the quantum phase
transition.
The only difference is that the effective field has different expression when
written as deviation from the field at $\nu=3/2$:
\be
B_{\rm eff} = \frac{B_\perp}{2p+1} = -3(B_\perp - B_{\perp,3/2}).
\ee
The extra factor of 3 comes from the fact that the number of holes change as
$B_\perp$ does, when the total number of electron is fixed.

What we have written until now have been known and have been used to analyze
the experiments.
Now we derive new relation by noticing that $\nu=2-p/(2p+1)$ can be written as
$\nu=1+p'/(2p'+1)$ with $p'=-p-1$.
If the Zeeman splitting is large enough, the filled down-spin Landau level is
inert and can be neglected.
Therefore in this situation the system is described by spinless CF at filling
factor $|p'|$.
What we want to do is to use this picture even if the Zeeman splitting becomes
smaller by considering the down-spin Landau level appropriately.

However, before taking into account the down-spin Landau level, a remark is in
order.
Namely we want to stress that the spin freedom of the CF comes from that of the
original electrons.
To see that lets assume that the CF has independent spin freedom.
Then we can consider that CF at $\nu_{\rm CF}=|p'|$ has Zeeman split spin
Landau levels which are empty.
When the Zeeman splitting is reduced, the quantum phase transition occurs
between these spin-split Landau levels.
The condition for the transition point is obtained by replacing $p$ in
eq.(\ref{cond1}) by $p'$, and is given by
\be
(|p'|-2k-1)\hbar\omega_{\rm c}^* =g^*\mu_{\rm B}B_{\rm tot}.
\ee
However, since $p'=-p-1$, this condition contradicts eq.(\ref{cond1}).
This clearly shows that the spin freedom of CF comes from that of the
electrons.
At the phase transition, electrons in the filled down-spin Landau level must
move to the up-spin Landau level.

Now let us describe the transition.
What we need to do is to establish a set of rules to treat the filled down-spin
Landau level at the composite fermion transformation.
The spirit of the composite fermion transformation is to make explicit the
already existing correlation between electrons by attaching the fictitious flux
quanta.
Since, only the minimal correlation due to the Fermi statistics is possible for
the filled Landau level, it does not seem appropriate to attach flux quanta to
the down-spin electrons.
However, the holes introduced in the down-spin Landau level repel each other
and they can have flux attached.
Another clue is that the effective magnetic field should not change
discontinuously at the quantum phase transition.
{}From these considerations we derive the following rules.
(1) All the electrons are changed into CF's.
(2) However, each single electronic state in the lowest down-spin Landau level
gives two flux quanta with opposite sign.
These two rules guarantee that the filled level does not contribute to
fictitious flux, and that the spin-flip of an electron will not change the
effective magnetic field.
The effective field is calculated as follows.
We consider a finite size system with $N_{\rm e}$ electrons and $N_0$
flux quantum.
Namely the degeneracy of a Landau level is $N_0$ and $N_{\rm e}=\nu N_0$.
The effective flux quanta is reduced from $N_0$ by $2N_{\rm e}$  and increased
by $2N_0$ according to the rules (1) and (2), respectively.
Thus
\begin{eqnarray}
N_{0,{\rm eff}} &=& N_0 - 2 N_{\rm e} +2 N_0 = (3 - 2\nu) N_0 \nonumber \\
 &=& \frac{1}{2p'+1} N_0.\\
\nonumber
\end{eqnarray}
Namely, the magnetic field is reduced by a factor of $(2p'+1)$:
$B_{\rm eff} = B_{\perp}/(|2p'+1|)=B_{\perp}/(|2p+1|)$.
Since the magnetic field is weaker by a factor of $|2p'+1|$ for the CF, the
down-spin CF fills $|2p'+1|$ Landau levels, when the down-spin Landau level is
filled by electrons.
As we have restricted ourselves to consider only the lowest Landau level, we
cannot accommodate more down-spin electrons.
This fact also gives an restriction to the CF filling factors, or an additional
rule, namely, (3) the maximum allowed CF filling factor for each spin state is
$|2p'+1|$.
Finally we impose the last rule: (4) the Zeeman splitting of these levels are
$g^*\mu_{\rm B} B_{\rm tot}$.

Now let us see these rules give correct result:
When the electronic filling factor is $\nu=1+p'/(2p'+1)$, the total CF filling
factor becomes $\nu_{\rm CF} = |2p'+1|+|p'|=|3p'+1|$.
The quantum phase transition occurs between the state where
\be
 \{
  \begin{array}{rl}
    \nu_{{\rm CF},\downarrow} &= |2p'+1|-k \\
    \nu_{{\rm CF},\uparrow} &= |p'| + k, \\
 \end{array}
\ee
to the state where
\be
 \{
  \begin{array}{rl}
    \nu_{{\rm CF},\downarrow} &= |2p'+1|-k-1 \\
    \nu_{{\rm CF},\uparrow} &= |p'|+ k+1, \\
 \end{array}
\ee
where $k$ is an integer.
As before from the condition that the energy of the down-spin HOLL coincide
with that of the up-spin LULL, we get the following equation,
\be
(|p'+1| -2k -1)\hbar\omega_{\rm c}^* =g^*\mu_{\rm B}B_{\rm tot}.
\ee
Since $|p'+1| =-p$, this condition is the same as eq.(\ref{cond1}).
I have considered possibilities for other rules.
However, all the plausible rules I coined could not give the transition point
consistent with eq.(\ref{cond1}).

%
%
\section{Rules and Application}

In this section we generalize the rules to the series of filling factors
related to $\nu=p/(mp+1)$, where $m$ is a positive even integer.
As we have seen above, states at $\nu=p/(mp+1)$ and at $\nu=2-p/(mp+1)$ are
simple: the states are transformed to the CF system at total filling factor
$|p|$.
The CF Landau levels have spacing $\hbar\omega_{\rm c}^* = \hbar eB_{\rm
eff}/m^*$, and they are spin-split by $g^*\mu_{\rm B}B_{\rm tot}$.

New series of filling factors for which we assign rules are
\be
\nu=1 + \frac{p}{mp+1},
\ee
and its electron-hole symmetric states at
\be
\nu=1 - \frac{p}{mp+1}.
\ee
These states are expressed by CFs at total filling factor
\be
\nu_{\rm CF}=|mp+1|+|p|,
\ee
according to the following rules:
(1) All the electrons are changed into CF's by attaching $m$ flux quanta.
(2) However, each electronic states in the lowest down spin Landau level gives
$|m|$ flux quanta with opposite sign.
(3) the maximum allowed CF filling factor for each spin state is $|mp+1|$.
(4) the Zeeman splitting of these levels are $g^*\mu_{\rm B} B_{\rm tot}$,
while the Landau splitting is given by the effective magnetic field,
\be
B_{\rm eff} = \frac{B_\perp}{mp+1}.
\ee
We can rewrite $B_{\rm eff}$ in a different way also.
It is proportional to the deviation of $B_\perp$ from that at $\nu=1 \pm 1/m$,
$B_{\perp,1\pm 1/m}$:
\be
B_{\rm eff} = \pm (m \pm 1) (B_\perp - B_{\perp,1\pm 1/m}).
\ee
The condition for the gap collapse, or the quantum phase transition point, is
given by these rules as follows:
\be
(|mp+1| - |p| -2k -1)\hbar\omega_{\rm c}^* =g^*\mu_{\rm B}B_{\rm tot},
\label{cond2}
\ee
where $k$ is a non-negative integer, and $\omega_{\rm c}^*=eB_{\rm eff}/m^*$.

Now that we have obtained the rules, we can analyze the case of $\nu=5/7$
mentioned in the introduction.
This filling factor belongs to a series $\nu = 1-p/(4p+1)$: $\nu=5/7$ is
obtained by putting $p=-2$.
Therefore, the transition occurs at
\be
(4-2k)\hbar\omega_{\rm c}^* = g^*\mu_{\rm B}B_{\rm tot}.
\ee
The transition at $k=1$ is clearly observed at the correct magnetic field by
Yeh {\it et al.}'s experiment as shown in Fig.2 in their paper.\cite{yeh}
The transition at $k=0$ is also expected to occur, but it requires stronger
magnetic field and has not been observed yet.

%
%
\section{Discussion}

So far experiments have been done around $\nu=3/2$\cite{du} and around
$\nu=3/4$.\cite{yeh}
The observed phase transition points are used to deduce the combination
$g^*m^*$.
Since, $\nu=3/2$ is electron-hole symmetric to $\nu=1/2$, previously known CF
theory is almost sufficient to explain the experiments around
$\nu=3/2$:\cite{du}
the longitudinal resistance shows distinct peaks where the gap is expected to
collapse.
Although, weak peaks are observed, where eq.(\ref{cond1}) is satisfied with
half-integer $k$.

For the interpretation of the experiment around $\nu=3/4$, we need the theory
in this paper.
Because, if we want to use the previously known simple theory, we must consider
the states around $\nu=3/4$ as electron-hole symmetric states of those around
$\nu=1/4$; in this case, the peak position does not coincide with the phase
transition point as remarked in the report of the experiment,\cite{yeh} and
stated in the introduction.
Comparison with the experiment shows that the present theory can correctly
explain the distinct peaks of the resistance at $\nu=8/11$ and 5/7.
These filling factors belong to negative $p$ side of the series
$\nu=1-p/(4p+1)$.
On the other hand, at $\nu=4/5$ and 7/9, namely positive $p$ side of the
series, the stronger peaks correspond to half-integer $k$.
This discrepancy and the existence of weaker peaks are left as problems to be
solved in the future.
It should be remarked that Wu {\it et al.}'s three step CF
transformation\cite{wu} also cannot resolve this discrepancy.
The existence of weaker peaks seems to indicate the CF theory is too naive,
even if the stronger peaks can be understood successfully.
Possible origin of the discrepancy at $\nu=4/5$ and 7/9 could be the exchange
enhancement of the Zeeman splitting or the effect of skyrmion.

The experimentally determined phase transition point gives the combination
$g^*m^*$ through the condition eq.(\ref{cond2}).
To get $g^*$ and $m^*$ separately, the experimentalists used the temperature
dependence of the Schubnikov-de Haas data to deduce $m^*$.
{}From the obtained values of $m^*$, $g^*$ of about 0.6 is obtained at
$\nu=5/7$ as well as around $\nu=3/2$.
They could not find a reason why the $g$-factor of the two-flux quanta CF is
equal to that of the four-flux quanta CF.
However, in the present simple theory, the $g$-factor of the CF is nothing but
that of original electrons, this coincidence is not strange at all.
Actually, 0.6 is close to the $g$-factor of electron, 0.44.

As stated in the introduction, a mapping of $\nu=5/7$ state to $\nu_{\rm CF}=1$
is possible.\cite{wu}
Namely, there are several schemes to map electronic states to CF states.
The present investigation clarified that the mappings are not equivalent:
in one of the mapping quantum phase transition never occurs, but it does in
another mapping.
Then how can we know which mapping is the correct one?
Of course experiments can discriminate them, but as a theoretical problem how
can we do that.
One criterion is whether the mapping is simple or not.
In the case considered in this paper our one-step mapping is much simpler than
three-step mapping by Wu {\it et al.}, and consistent with the experiment.
``Simple is best" is comfortable for physicists, but we need to give foundation
for this conjecture, which is left as a future problem.

In conclusion in this paper we developed a set of rules for the CF
transformation, which can explain the quantum phase transition observed
experimentally.
We also found that there still remain several problems to be solved in the
future.

\section*{Acknowledgements}
I thank Dan Tsui who showed me the experimental results prior to publication
while we stayed at Aspen Center for Physics, where part of this work was done.
This work is supported by Grant-in-Aid for Scientific Research (C) 10640301
from the Ministry of Education, Science, Sports and Culture.


\begin{references}
\bibitem{klitz} K. von Klitzing, G. Dorda and M. Pepper: Phys. Rev. Lett.
{\bf 45} (1980) 494.
\bibitem{prange} {\it The Quantum Hall Effect}, ed. R.E. Prange and
S.M. Girvin (Springer, New York, 1990) 2nd ed.
\bibitem{jain} J.K. Jain: Phys. Rev. Lett. {\bf 63} (1989) 199.
\bibitem{heinonen} {\it Composite Fermions} ed. O. Heinonen
(World Scientific, Singapore, 1998).
\bibitem{du} R.R. Du, A.S. Yeh, H.L. Stormer, D.C. Tsui, L.N. Pfeiffer
and K. West: Phys. Rev. Lett. {\bf 75} (1995) 3926.
\bibitem{yeh} A.S. Yeh, H.L. Stormer, D.C. Tsui, L.N. Pfeiffer,
K.W. Baldwin
and K. West: Phys. Rev. Lett. {\bf 82} (1999) 592.
\bibitem{wu} X.G. Wu, G. Dev and J.K. Jain: Phys. Rev. Lett. {\bf 71}
(1993) 153.
\bibitem{jain1} It was pointed out by a referee to ref.(\cite{dy})
that $\nu=5/7$ is transformed in this way.
\bibitem{dy} D. Yoshioka: Phys. Rev. Lett. {\bf 83} (1999) 886.
\end{references}
\end{document}